\documentclass[aps, showpacs, amsmath, dvips, showkeys, preprint, nofootinbib]{revtex4}

\usepackage{graphicx}   
\usepackage{color}      
\usepackage{braket}
\usepackage{tikz}
\usepackage{pgfplots}
\def \be  {\begin{equation}}
\def \ee  {\end{equation}}
\def \ee  {\end{equation}}
\def \bea {\begin{eqnarray}}
\def \eea {\end{eqnarray}}

\newcommand{\nn}{\nonumber}

\begin{document}

\preprint{ECTP-2020-05}
\preprint{WLCAPP-2020-05}
\hspace{0.01cm}

\title{A Possible Solution of the Cosmological Constant Problem based on Minimal Length Uncertainty and GW170817 and PLANCK Observations}

\author{Abdel Magied Diab}
\email{a.diab@eng.mti.edu.eg}
\affiliation{Modern University for Technology and Information (MTI), Faculty of Engineering, 11571 Cairo, Egypt}

\author{Abdel Nasser Tawfik} 
\email{tawfik@nu.edu.eg}
\affiliation{Nile University, Egyptian Center for Theoretical Physics (ECTP), Juhayna Square of 26th-July-Corridor, 12588 Giza, Egypt}

\date{\today}

\begin{abstract}

We propose the generalized uncertainty principle (GUP) with an additional term of quadratic momentum motivated by string theory and black hole physics as a quantum mechanical framework for the minimal length uncertainty at the Planck scale. We  demonstrate that the GUP parameter, $\beta_0$, could be best constrained by the the gravitational waves observations; GW170817 event. Also, we suggest another proposal based on the modified dispersion relations (MDRs) in order to calculate the difference between the group velocity of gravitons and that of photons. We conclude that the upper bound reads $\beta_0 \simeq 10^{60}$. Utilizing features of the UV/IR correspondence and the obvious similarities between GUP (including non-gravitating and gravitating impacts on Heisenberg uncertainty principle) and the discrepancy between the theoretical and the observed cosmological constant $\Lambda$ (apparently manifesting gravitational influences on the vacuum energy density), known as {\it catastrophe of non-gravitating vacuum}, we suggest a possible solution for this long-standing physical problem, $\Lambda \simeq 10^{-47}~$GeV$^4/\hbar^3 c^3$.

\end{abstract}


\maketitle

\makeatletter
\let\toc@pre\relax
\let\toc@post\relax
\makeatother 


\section{Introduction} 
\label{intro}

The cosmological constant, $\Lambda$, an essential ingredient of the theory of general relativity (GR) \cite{Einstein1917As}, was guided by the idea that the evolution of the Universe should be static \cite{Tawfik:2011mw,Tawfik:2008cd}. This model was subsequently refuted and accordingly the $\Lambda$-term was abandoned from the Einstein field equation (EFE), especially after the confirmation of the celebrated Hubble obervations in 1929 \cite{Hubble:1929ig}, which also have verified the consequences of Friedmann solutions for EFE with vanishing $\Lambda$ \cite{Friedman:1922kd}. Nearly immediate after publishing GR, a matter-free solution for EFE with finite $\Lambda$-term was obtained by de Sitter  \cite{deSitter:1917zz}. Later on when it has been realised that the Einstein {\it static} Universe was found unstable for small perturbations \cite{Mulryne:2005ef, Wu:2009ah, delCampo:2011mq}, it was argued that the inclusion of the $\Lambda$-term remarkably contributes to the stability and simultaniously supports the expansion of the Universe, especially that the initial singularity of Friedmann-Lem$\hat{\mbox{a}}$itre-Robertson-Walker (FLRW) models could be improved, as well \cite{Weinberg1972AA,Misner1984B}. Furthermore, the observations of type-Ia high redshift supernovae in late ninteeth of the last century \cite{Riess:1998cb, Perlmutter:1998np} indicated that the expanding Universe is also accelerating, especially at a small $\Lambda$-value, which obviously contributes to the cosmic negative pressure \cite{Garriga:1999bf,Martel:1997vi}. With this regard, we recall that the cosmological constant can be related to the vacuum energy density, $\rho$, as $\Lambda=8\pi G \rho/c^2$, where $c$ is the speed of light in vacuum and $G$ is the gravitational constant. In 2018, the PLANCK observations have provided us with a precise estimation of $\Lambda$, namely $\Lambda_{\mbox{Planck}} \simeq 10^{-47}$GeV$^4/\hbar^3 c^3$ \cite{Aghanim:2018eyx}. When comparing this tiny value with the theoretical estimation based on quantum field theory in weakly- or non-gravitating vacuum, $\Lambda_{\mbox{QFT}} \simeq 10^{74}$GeV$^4/\hbar^3 c^3$, there is, at least, a $121$-orders-of-magnitude-difference to be fixed \cite{Adler:1995vd,Weinberg:1988cp,Zeldovich:1968ehl}.

The disagreement between both values is one of the greatest mysteries in physics and known as the cosmological constant problem or {\it catastrophe of non-gravitating vacuum}. Here, we present an attempt to solve this problem. To this end, we utilize the generalized uncertainty principle (GUP), which is an extended version of Heisenberg uncertainty principle (HUP), where a correction term encompassing the gravitational impacts is added, and thus an alternative quantum gravity approach emerges \cite{Tawfik:2014zca,Tawfik:2015rva}. To summarize, the present attempt is motivated by the similarity of GUP (including non-gravitating and gravitating impacts on HUP) and the disagreement between theoretical and observed estimations for $\Lambda$ (manifesting gravitational influences on the vacuum energy density) and by the remarkable impacts of $\Lambda$ on early and late evolution of the Universe \cite{Tawfik:2019jsa,Tawfik:2011mw,Tawfik:2008cd}. So far, there are various quantum gravity approaches presenting quantum descriptions for different physical phenomena in presence of gravitational fields to be achnowledged, here \cite{Tawfik:2014zca, Tawfik:2015rva}. 

The GUP offers a quantum mechanical framework for a potential minimal length uncertainty in terms of the Planck scale \cite{Tawfik:2017syy,Tawfik:2016uhs,Dahab:2014tda,Ali:2013ma}. The minimal length uncertainty, as proposed by GUP, exhibits some features of the UV/IR correspondence \cite{Maldacena:1997re,Gubser:1998bc,Witten:1998qj}, which has been performed in viewpoint of local quantum field theory. Thus, it is argued that the UV/IR correspondence is relevant to revealing several aspects of short-distance physics, such as, the cosmological constant problem \cite{Weinberg:1988cp,Banks:2000fe,Cohen:1998zx,ArkaniHamed:2000eg}. Therefore, a precise estimation of the minimal length uncertainty strongly depends on the proposed upper bound of the GUP parameter, $\beta_0$ \cite{Dahab:2014tda,Tawfik:2013uza}.

Various ratings for the upper bound of $\beta_0$ have been proposed, for example, by comparing quantum gravity corrections to various quantum phenomena with electroweak  \cite{Das:2008kaa, Das:2009hs} and astronomical \cite{ Scardigli:2014qka,  Feng:2016tyt} observations. Accordingly, $\beta_0$ ranges between $10^{33}$ to $10^{78}$ \cite{Scardigli:2014qka,Feng:2016tyt,Walker:2018muw}. As a preamble of the present study, we present a novel estimation for $\beta_0$ from the binary neutron stars merger, the gravitational wave event GW170817 reported by the Laser Interferometer Gravitational-Wave Observatory (LIGO) and the Advanced Virgo collaborations \cite{TheLIGOScientific:2017qsa}. With this regard, there are different efforts based on the features of the UV/IR correspondence in order to interpret the $\Lambda$ problem \cite{Chang:2001bm, Chang:2011jj,Miao:2013wua,Shababi:2017zrt,Vagenas:2019wzd} with Liouville theorem in the classical limit \cite{Fityo:2008zz, Chang:2001bm, Wang:2010ct}. Having a novel estimation of $\beta_0$, a solution of the $\Lambda$ problem, {\it catastrophe of non-gravitating vacuum}, could be best proposed. 

The present paper is organized as follows. Section \ref{MDRGUP} reviews the basic concepts of the GUP approach with quadratic momentum. The associated modifications of the energy-momentum dispersion relations related to GR and rainbow gravity are also outlined in this section. In section \ref{GUPparameter}, we show that the dimensionless GUP parameter, $\beta_o$, could be, for instance, constrained to the gravitational wave event GW170817. Section \ref{LamdaProblem} is devoted to calculating the vacuum energy density of states and shows how this contributes to understanding the cosmological constant problem with an quantum gravity approach, the GUP. The final conclusions are outlined in section \ref{conclusion}.

\section{Generalized Uncertainty Principle and Modified Dispersion Relations \label{MDRGUP}}

Several approaches to the quantum gravity, such as GUP, predict a minimal length uncertainties that could be related to the Planck scale \cite{Tawfik:2015rva,Tawfik:2014zca}. There were various laboratory experiments conducted to examine the GUP effects \cite{Bawaj:2014cda,  Marin:2013pga, Pikovski:2011zk, Khodadi:2018kqp}. In this section, we focus the discussion on GUP with a quadratic momentum uncertainty \cite{Tawfik:2015rva,Tawfik:2014zca}. This version of GUP was obtained from black hole physics \cite{Gross:1987kza} and supported by {\it gedanken} experiments \cite{Maggiore:1993zu}, which have been proposed Kempf, Mangano, and Mann (KMM), \cite{Kempf:1994su} 
\bea
\Delta x\, \Delta p\geq \frac{\hbar}{2} \left[ 1+ \beta (\Delta p)^2 \right], \label{GUPuncertainty}
\eea 
 where $\Delta x$ and $\Delta p$ are the uncertainties in position and momentum, respectively.  The GUP parameter can be exressed as $\beta = \beta_0 (\ell_p/\hbar)^2 = \beta_0/ (M_p c)^2$, where $\beta_0$ is a dimensionless parameter, $\ell_p=1.977 \times 10^{-16}~$GeV$^{-1}$ is the Planck length, and $M_p= 1.22 \times 10^{19}~$GeV$/c^2$ is the Planck mass. Equation (\ref{GUPuncertainty}) implies the existence of a minimum length uncertainty, which is related to the Planck scale, $\Delta x_{\mbox{min}} \approx \hbar \sqrt{\beta} =\ell_p \sqrt{\beta_0}$. It should be noticed that the minimum length uncertainty exhibits features of the UV/IR correspondence \cite{Maldacena:1997re,Gubser:1998bc,Witten:1998qj}. $\Delta x$ is obviously proportional to $\Delta p$, where large $\Delta p$ (UV) becomes proportional to large  $\Delta x$ (IR).  Equation (\ref{GUPuncertainty}) is a noncommutative relation; $[\hat{x}_i,\; \hat{p}_j] = \delta_{ij} i \hbar [1+\beta p^2]$, where both position and momentum operators can be defined as  
\bea 
 \hat{x}_i = \hat{x}_{0i},  \quad \quad
 \hat{p}_j= \hat{p}_{0j} (1+\beta p^2),
 \eea
where $\hat{x}_{0i}$ and $\hat{p}_{0j}$ are corresponding operators obtained from the canonical commutation relations  $[\hat{x}_{0i},\; \hat{p}_{0j}]=\delta_{ij} i \hbar,$ and $p^2= g_{ij} p^{0i} \; p^{0j}$. 

We can now construct the modified dispersion relation (MDR) due to quadratic GUP. We start with the background metric in GR gravitational spacetime
\bea
ds^2 =g_{\mu \nu} dx^\mu \,  dx^\nu = g_{00} c^2 dt^2 +  g_{ij} dx^i\,  dx^j,
\eea  
with $g_{\mu \nu}$ is the Minkowski spacetime metric tensor $(-,+,+,+)$. Accordingly, the modified four-momentum squared is given by 
\bea
p_\mu p^\mu = g_{\mu \mu} p^\mu p^\mu  &=& g_{00} (p^0)^2 + g_{ij}  p^{0i}  p^{0j} (1+\beta p^2)  \nn  \\ 
&=& -(p^0)^2  + p^2 + 2 \beta\; p^2 \,\cdot\, p^2.  \label{modifyMomentum}
\eea
Comparing this with the conventional dispersion relation, $p_\mu p^\mu = - m^2c^2$, the time component of the momentum can then be written as 
\bea
(p^0)^2 &=& m^2c^2 + p^2  (1+\beta p^2).
\eea
The energy of the particle $\omega$ can be defined as $\omega/c = - \zeta_\mu p^\mu = - g_{\mu \nu} \zeta^\mu p^\nu$, where the killing vector is given as $\zeta^\mu = (1,0,0,0)$. Therefore, the energy of the particle could be expressed as $\omega=-g_{00} c (p^0)=c (p^0)$ and the modified dispersion relation in GR gravity reads
\bea
\omega^2 = m^2 c^4 + p^2 c^2 (1+ 2\beta p^2). \qquad\qquad\qquad\mbox{GR Gravity} \label{MDRrel} 
\eea
For $\beta \rightarrow 0$, the standard dispersion can be obtained. 

The rainbow gravity generalizes the MDR in doubly special relativity to curved spacetime \cite{magueijo2004gravity}, where the geometry spacetime is explored by a test particle with energy $\omega$ \cite{Magueijo:2001cr, Magueijo:2002am},
\bea
\omega^2 \; f_1 \left(\frac{\omega}{\omega_p}\right)^2 - (pc)^2 f_2 \left(\frac{\omega}{\omega_p}\right)^2 = \left(mc^2\right)^2,
\eea
where $\omega_p$ is the Planck energy and $f_1 (\omega/\omega_p)$ and $f_2 (\omega/\omega_p)$ are known as the rainbow functions which are model-depending. The rainbow functions can be defined as \cite{AmelinoCamelia:1996pj, AmelinoCamelia:1997gz},
\bea
f_1 (\omega/\omega_p) &=& 1, \quad \quad f_2 (\omega/\omega_p) = \sqrt{1-\eta   (\omega/\omega_p)^n}, \label{Rainbowfuncs}
\eea
where $\eta$ and $n$ are free positive parameters. It was argued that for the logarithmic corrections of black hole entropy \cite{Tawfik:2015kga}, the integer $n$ is limited as $n=1,2$ \cite{Gangopadhyay:2016rpl}. Therefore, it would be eligible to assume that $n=2$. Thus, the MDR for rainbow gravity with GUP can be written as,
\bea
\omega^2 = \frac{(mc^2)^2  +  p^2 c^2 (1+2\beta p^2)}{1+ \eta\; \Big[\frac{pc}{\omega_p}\Big]^2 (1+2\beta p^2)}.\qquad\qquad\qquad\mbox{Rainbow Gravity}   \label{MDRrain}
\eea
Again, as $\beta \rightarrow 0$, Eq. (\ref{MDRrain}) goes back to the standard dispersion relation. 

We have constructed two different MDRs for quadratic GUP, namely Eqs (\ref{MDRrel}) and (\ref{MDRrain}) in GR and rainbow gravity, respectively. Bounds on GUP parameter from GW170817 shall be outlined in the section that follows.

\section{Bounds on GUP parameter from GW170817}
\label{GUPparameter}

Instead of violating Lorentz invariance \cite{Tawfik:2012hz}, we intend to investigate the speed of the graviton from the GW170817 event. To this end, we use MDRs obtained from the quadratic GUP approaches, section \ref{MDRGUP}. Thus, defining an upper bound on the dimensionless GUP parameter $\beta_0$ for given bounds on mass and energy of the graviton, where $m_g\lesssim 4.4 \times 10^{-22}~$eV$/c^2$ and $\omega = 8.5 \times 10^{-13}~$eV, respectively, plays an essential role. Assuming that the gravitational waves propagate as free waves, we could, therefore, determine the speed of the mediator, that of the graviton, from the group velocity of the accompanying wavefront, i.e. $v_g = \partial \omega/\partial p$, where $\omega$ and $p$ are the energy and momentum of the graviton, respectively \cite{Mirshekari:2011yq}. The idea is that the group velocity of the graviton can be simply deduced from the MDRs, Eqs. (\ref{MDRrel}) for the GR gravity and (\ref{MDRrain}) and the rainbow gravity, in presence and then in absence of the GUP impacts, which have been discussed in section \ref{MDRGUP}. Accordingly, Eq. (\ref{MDRrel}) implies that the group velocity reads
\bea
v_g = \frac{\partial \omega}{\partial p} = \frac{pc^2}{\omega} \left(1+ 4 \beta p^2\right). \label{vgMDR1}
\eea 

The unmodified momentum $p$ in terms of the modified parameters up to $\mathcal{O} (\beta)$, can be expressed as $p=a+b \beta$, where $a$ and $b$ are arbitrary parameters. By substituting this expression into Eq. (\ref{MDRrel}), we find that $p^2=  (\omega_g/c)^2 - m^2 c^2 $. Thus,  Eq. (\ref{vgMDR1}) can be rewritten as
\bea
v_g = c \;\Big\{ \Big[1-\Big( \frac{mc^2}{\omega_g}\Big)^2 \Big]^{1/2} +4\beta \frac{\omega_g^2}{c^2} \; \Big[1-\Big( \frac{mc^2}{\omega_g}\Big)^2 \Big]^{3/2} \Big\}, 
\eea
where $\omega_g$ is the energy of the graviton. It is obvious that for $\beta \rightarrow 0 $, i.e. in absence of GUP impacts, the group velocity reads
\bea
v_g  = c \Big[1- \frac{1}{2}\Big( \frac{mc^2}{\omega_g}\Big)^2 \Big].
\eea  
Then, the difference between the speed of photon (light) and that of graviton without GUP impacts is given as 
\bea
\Big| \delta v\Big| = \Big| c-v_g\Big|  = c \Big| \frac{1}{2}\Big( \frac{mc^2}{\omega_g}\Big)^2 \Big| \lesssim 1.34 \times 10^{-19}\;c. \label{vDr}
\eea
Although the small difference obtained, we are - in the gravitational waves epoch - technically able to measure even a such tiny difference! In light of this, we could use the results associated with the GW170817 event, such as the graviton velocity, in order to set an upper bound on the GUP parameter, $\beta_0$. 

For a massless graviton, the difference between the speed of photons (light) and that of the gravitons in presence of the GUP impacts reads 
\bea
\Big|\delta v_{\mbox{GUP}}\Big| &=&\Big| 4\beta \frac{\omega^2}{c}\Big|  =   \Big| 4\beta_0 \Big(\frac{\omega^2}{ M_p^2 c^3}\Big)^2\Big| \lesssim 1.95 \times 10 ^{-80} \beta_0 \;c.     \label{vMDR}
\eea 
Thus, the upper bound on the dimensionless parameter, $\beta_0$, of the quadratic GUP can be simply deduced from Eqs. (\ref{vDr}) and (\ref{vMDR}), 
\bea
\beta_0 \lesssim 8.89 \times 10^{60}. \label{MDRbeta1}
\eea  

The group velocity of the graviton due to MDR and rainbow gravity when applying the quadratic GUP approach, Eq. (\ref{MDRrain}), can be expressed as
\bea
v_g = \frac{\partial \omega}{\partial p} = \Big(\frac{pc^2}{\omega_g}\Big) \; \frac{\Big( 1- \frac{\eta}{\omega_p^2} (mc^2)^2  \Big) \Big(1+4\beta p^2 \Big)}{\left[1+ \eta \Big(\frac{cp}{\omega_p}\Big)^2 (1+ 2 \beta p^2) \right]^2}.
\eea
Similarly, one can for a massless graviton express the conventional momentum in terms of the GUP parameter. In order of $\mathcal{O}(\beta)$, we get  
\bea
c p = \omega_g  \Big[ \Big(1 - \eta \Big(\frac{\omega_g}{\omega_p}\Big)^2  \Big)^{-1/2} - \beta
\frac{\omega_g^2}{c^2} \Big(1 -\eta \Big(\frac{\omega_g}{\omega_p}\Big)^2  \Big)^{-3/2} \Big]. \label{pcRainbow}
\eea   
The unmodified momentum can be expressed in GUP-terms up to $\mathcal{O}(\beta)$; $p=a_0 + a_1 \beta$, where $a_0$ and $a_1$ are arbitrary parameters. Nevertheless, the investigation of the speed of the graviton from the GW150914 observations \cite{Abbott:2016blz} specifies the rainbow gravity parameter, $ \eta (\omega_g/\omega_p)^2\leq 3.3\times 10^{-21}$  \cite{Gwak:2016wmg}. Accordingly, Eq. (\ref{pcRainbow}) can be reduced to $c p=\omega_g (1- \beta \omega_g^2/c^2)$ and the group velocity of the massless graviton becomes 
\bea
v_g = c \Big[ 1 - 5\frac{\beta \omega^2}{c^2} + \mathcal{O}(\beta^2) \Big].
\eea

Then, the difference between the speed of photons and that of the gravitons reads
\bea
\Big|\delta v_{\mbox{GUP}}\Big| &=&\Big| 5 \beta \frac{\omega^2}{c} \Big| \lesssim 2.43 \times 10^{-80} \beta_0 \,  c.  \label{vgRainbow}
\eea
By comparing  Eqs. (\ref{vgRainbow}) and (\ref{vDr}), the upper bound of the GUP parameter $\beta_0$ can be estimated as
\bea
\beta_0 \lesssim 5.5 \times 10^{60}.  \label{Rainbowbeta1}
\eea

It is obvious that both results, Eqs. (\ref{MDRbeta1}) and (\ref{Rainbowbeta1}), are very close to each other; $\beta_0 \lesssim 10^{60}$. The improved upper bound of $\beta_0$ is very similar to the ones reported in refs.  \cite{Scardigli:2014qka, Feng:2016tyt}, which - as well -  are depending on astronomical observations. The present results are based on mergers of spinning neutron stars. Thus, it is believed that more accurate observations, the more precise shall be $\beta_0$. 

Having set a upper bound on the GUP parameter and counting on the spoken similarities between GUP and the catastrophe of non-gravitating vacuum, we can now propose a possible solution of the cosmological constant problem.

\section{A Possible Solution of the Cosmological Constant Problem} 
\label{LamdaProblem}

The cosmological constant can be given as $\Lambda = 3 H_0^2 \Omega_\Lambda$, where $H_0$ and $\Omega_\Lambda$ are the Hubble parameter and the dark energy density, respectively \cite{Carroll:2000fy}. On the other hand, the origin of the catastrophe of non-gravitating vacuum would be understood from the disproportion of the value of $\Lambda$ in the theoretical calculations, while this is apparently impacting the GW observations \cite{Sahni:2002kh}. From the most updated PLANCK observations, the values of $\Omega_\Lambda = 0.6889 \pm 0.0056$ and $H_0 = 67.66 \pm 0.42~$Km $\cdot$ s$^{-1}$ $\cdot$  Mpc$^{-1}$ \cite{Aghanim:2018eyx}. Then, the vacuum energy density
\bea
\frac{c^2}{8 \pi G} \Lambda &=& \left(\frac{3 H_0^2 c^2}{8\pi G}\right) \Omega_\Lambda = \frac{3\hbar c}{8\pi \ell_p^2 \ell_0^2} \Omega_\Lambda,  \label{VacuEnergy}
\eea
where the scale of the visible light, $\ell_0= c/H_0 \simeq 1.368 \times 10^{23}~$Km \cite{Aghanim:2018eyx}. Therefore, one can use Eq. (\ref{VacuEnergy}) to esiamte the vacuum energy density in order of $10^{-47}~$GeV$^4/(\hbar^3c^3)$. In quantum field theory, the cosmological constant is to be calculated from sum over the vacuum fluctuation energies corresponding to all particle momentum states \cite{Carroll:2000fy}. For a massless particle, we obtain 
\bea
\frac{1}{(2\pi \hbar)^3} \int d^3 \; \vec{p}\; (\hbar \omega_p /2) \simeq 9.60\times10^{74} \; {\mbox{GeV}}^4/ (\hbar^3 c^3).  \label{QFTlamda}
\eea
This is clearly infinite integral. But, it is usually cut off, at the Planck scale, $\mu_p = \hbar/\ell_p$. We assume $\omega_p$ is the vacuum energy of quantum harmonic state $\hbar \omega_p = [p^2c^2+m_g^2c^4]^{1/2}$.  

To propose a possible solution of the cosmological constant problem, it is initially needed to determine the number of states in the phase space volume taking into account GUP, Eq. (\ref{GUPuncertainty}). An analogy can be found in Liouville theorem in the classical limit. We need to make sure that the size of each quantum mechanical state in phase space volume is depending on the modified momentum $p$, especially when taking GUP into consideration, Eq. (\ref{GUPuncertainty}). In other words, the number of quantum states in the phase space volume is assumed not depending on time. 

In the classical limit, the relation of the quantum commutation relations and the Poisson brackets is given as $[\hat{A}, \hat{B}] = i\hbar \{A, B\}$. Details on the Poisson bracket in D-Dimensions are outlined in appendix \ref{LiouvilleTheorem}.
Consequently, the modified density of states implies different implications on quantum field theory, such as, the cosmological constant problem. 

In D-dimensional spherical coordinate systems, the density of states in momentum space is given as \cite{Fityo:2008zz, Chang:2001bm, Wang:2010ct}   
\bea 
\frac{V\, d^D  \vec{p}}{(1+\beta p^2)^{D} },
\eea
where $V$ is the volume of space. It should be noticed that in quantum mechanics, the number of quantum stated per unit volume is given as $V/(2\pi \hbar)^D$. Therefore, for Liouville theorem, the weight factor in 3-D dimension reads \cite{Fityo:2008zz, Chang:2001bm, Wang:2010ct} (review appendix \ref{LiouvilleTheorem}) 
\bea
\frac{1}{(2\pi \hbar)^3} \frac{d^3 \vec{p}}{(1+\beta p^2)^3}. \label{densityStates}
\eea    
In quantum field theory, the modification in the quantum number of state of the phase space volume should have consequences on different quantum phenomena, such as, the  cosmological constant problem and the black body radiation. At finite weight factor of GUP, the sum over all momentum states per unit volume of the phase space modifies the vacuum energy density. The cosmological constant, on the other hand, is determined by summing over the vacuum fluctuations, the energies, corresponding to a particular momentum state
\bea
\Lambda_{\mbox{GUP}} (m) &=& \frac{1}{(2\pi \hbar)^3} \int  d^3 \vec{p} \rho(p^2) (\hbar \omega_p /2) = \frac{1}{2(2\pi \hbar)^3} \int   \frac{d^3 \vec{p}}{(1+\beta p^2)^3} \sqrt{p^2c^2+m_g^2c^4} 
\eea
For a massless particle, the vacuum energy density, which is directly related to $\Lambda$, reads
\bea
\Lambda_{\mbox{GUP}}(m=0) &=& \ \frac{c}{4\pi^2 \hbar^3} \int \frac{p^3}{(1+\beta p^2)^3}\; dp =   \frac{c (M_p^2 c^2)^2}{16 \pi^2 \hbar^3 \beta_0^2} = 1.78 \times 10^{-48}~\mbox{GeV}^4/(\hbar^3 c^3). \label{GUPLamda}
\eea
The agreement between the observed value of the cosmological constant, $\Lambda \simeq 10^{-47}~$GeV$^4/\hbar^3 c^3$, and our calculations based on quantum gravity approach, Eq. (\ref{GUPLamda}), is very convincing. We conclude that the connection between the estimated upper bound on $\beta_0$,  Eqs. (\ref{vgRainbow}) and (\ref{vDr}), from GW170817 event \cite{TheLIGOScientific:2017qsa} and the most updated observations of the PLANCK collaboration \cite{Aghanim:2018eyx} for the cosmological constant  $\Lambda$, Eq. (\ref{QFTlamda}), and our estimated value of $\Lambda(m=0)$, Eq. (\ref{GUPLamda}), gives an interpretation for the cosmological constant problem in presence of the minimal length uncertainty.

\section{Conclusions \label{conclusion} }     

In the present study, we have proposed the generalized uncertainty principle (GUP) with an addition term of quadratic momentum, from which we have driven the modified dispersion relations for GR and rainbow gravity, Eq. (\ref{MDRrel}) and Eq. (\ref{MDRrain}), respectively. Counting on the similarities between GUP (manifesting gravitational impacts on HUP) and the likely origin of the great discrepancy between the theoretical and observed values of the cosmological constant that in the gravitational impacts on the vacuum energy density, the present study suggests a possible solution for the long-standing cosmological constant problem ({\it catastrophe of non-gravitating vacuum}) that $\Lambda \simeq 10^{-47}~$GeV$^4/\hbar^3 c^3$.

We have assumed that the gravitational waves propagate as a free wave. Therefore, we could drive the group velocity in terms of the GUP parameter $\beta_0$ for GR and rainbow gravity, Eq. (\ref{MDRbeta1}) and Eq. (\ref{Rainbowbeta1}), respectively. Moreover, we have used recent results on gravitational waves, the binary neutron stars merger, GW170817 event, in order to determine the speed of the gravitons. Then, we have calculated the difference between the speed of gravitons and that of (photons) light, at finite and visnishing GUP parameter. We have shown that the upper bound on the dimensionless GUP parameter, $\beta \sim 10^{60}$, is merely constrained by such a speed difference. We have concluded that the speed of graviton is directly related to the GUP approach utilized in.  
 
The cosmological constant problem, which is stemming from the large discrepancy between the QFT-based calculations and the cosmological observations, is tagged as $\Lambda_{QFT}/\Lambda_{exp} \sim 10^{121}$. This quite large ratio can be interpreted by features of the UV/IR correspondence and the impacts of gravity. For the earlier, the large $\Delta x$ (IR) corresponds to a large $\Delta p$ (UV) in scale of Planck momentum. For the later, the GUP approach, for instance, Eq. (\ref{GUPuncertainty}), plays an essential role. We have assumed that in calculating the density of states where GUP approach is taken into account, a possible solution of the cosmological constant problem, Eq. (\ref{densityStates}), can be proposed. At Planck scale, the resulting density of the states seems to impact the vacuum energy density of each quantum state, Eq. (\ref{GUPLamda}). A refined value of the cosmological constant we have obtained for a novel upper bound on $\beta_0$, which - in turn - was determined from the GW170817 observations. Finally, the possible matching between the estimation of the upper bound on the GUP parameter deduced from the gravitational waves, GW170817 event, and the one estimated from the PLANCK 2018 observations seems to support the conclusion about the great importance of constructing a theory for quantum gravity. This likely helps in explaining various still-mysterious phenomena in physics.

\appendix
\section{Algebra of quantum mechanical commutators and Poisson brackets \label{LiouvilleTheorem}}

For a binary set of anticommutative functions on position and momentum, for instance, in D-dimensions, the Poisson bracket expresses their binary operation
\bea
\left\{ F(x_1, \cdots x_D;\; p_1, \cdots p_D  ),  G(x_1, \cdots x_D;\; p_1, \cdots p_D  )\right\} &=&  \nn\\
\left(\frac{\partial F}{\partial x_i} \, \frac{\partial G}{\partial p_j} - \frac{\partial F}{\partial p_i} \frac{\partial G}{\partial x_j} \right) \left\{ x_i, p_j \right\} &+& \frac{\partial F}{\partial x_i} \frac{\partial G}{\partial x_j} \left\{ x_i, x_j \right\}. 
\eea   
During a time duration, $\delta t$, the Hamilton's equations of motion for position and momentum can be given as 
\bea
x_i^\prime = x_i + \delta x_i,   \qquad \qquad  p_i^\prime = p_i + \delta p_i,
\eea
where, 
\bea 
 \delta x_i,    &=& \{ x_i, H\} \delta t = \{ x_i, p_j \}  \frac{\partial H}{\partial p_j} + \{ x_i, x_j\} \frac{H}{xj}, \\ 
 \delta p_i,    &=& \{ p_i, H\} \delta t = - \{ x_i, p_j \}  \frac{\partial H}{\partial x_j},
 \eea
where $H\equiv H(x,p;t)$ is the Hamiltonian, itself.

The estimation of the change in the phase space volume during the time evolution requires to determine the Jacobain of the transformation from $(x_1, \cdots x_D;\; p_1, \cdots p_D)$ to $(x_1^\prime, \cdots x_D^\prime;\; p_1^\prime, \cdots p_D^\prime)$, i.e.
\bea
d^Dx^\prime\; d^D p^\prime = \frac{d^Dx\; d^D p}{\mathcal{J}},
\eea
where $\mathcal{J}$ is the Jacobain of the transformation, which can be expressed as 
\bea
\mathcal{J} &=& \Big\| \frac{\partial (x_1^\prime, \cdots x_D^\prime;\; p_1^\prime, \cdots p_D^\prime  )}{\partial (x_1, \cdots x_D;\; p_1, \cdots p_D  ) } \Big\|  = 
1 + \left( \frac{\partial}{\partial x_i} \frac{\partial(\delta x_i)}{\partial t} + \frac{\partial}{\partial p_i} \frac{\partial(\delta p_i)}{\partial t} \right) \times \delta t. 
\eea
The general notations of position and momentum brackets lead to following algebraic relations 
\bea
\big\{x_i p_i\big\} = f_{ij} (x, p), \qquad  \big\{x_i, x_j\big\} = g_{ij}(x,p),\; \qquad  \mbox{and} \qquad \big\{p_i,p_j\big\}= h_{ij}(p). 
\eea
Thus,  the Jacobain of the transformation is given as \cite{Fityo:2008zz}
\bea 
\mathcal{J} = \prod_{i=1}^D f_{ii}(x,p)  =  1+ \sum_{i=1}^D (f_{ii} (x,p) - 1). \label{jacobian}
\eea
Therefore the invariant phase space in D-dimension reads
\bea
\frac{d^Dx d^D p}{(1+\beta p^2)^D}.
\eea  
Finally, the quantum density of states can be determined from  
\bea
\frac{1}{(2\pi \hbar)^3} \frac{d^3 \vec{p}}{(1+\beta p^2)^3}.
\eea  
%
\bibliographystyle{aip}
\bibliography{upperBoundofGUPParameterV1}

\end{document}